# Improvement of photosynthetic rate evaluation by plant bioelectric potential using illuminating information and a neural network

Ki Ando[a,*], Hiroshi Igarashi[a], Hiroyuki Shinoda[a], and Nobuki Mutsukura[a]

[a]Tokyo Denki University, 5 Senjyu-Asahi-cho, Adachi-ku, Tokyo 120-8551 Japan
*Corresponding author    E-mail: kiando@mail.dendai.ac.jp (K. Ando)

**Abstract**

The plant bioelectric potential is believed to be a suitable real-time and noninvasive method that can be used to evaluate plant activities, such as the photosynthetic reaction. The amplitude of the bioelectric potential response when plants are illuminated is correlated with the photosynthetic rate. However, practically, the bioelectric potential is affected by various cultivation parameters. This study analyzes the relationship between the bioelectric potential response and the illuminating parameters using a neural network to improve the accuracy of the photosynthetic rate evaluation. The variation of the illuminating colors to the plant affected the relationship between the amplitude of the bioelectric potential response and the photosynthetic rate; therefore, evaluating the photosynthetic rate using the amplitude is difficult. The analysis result shows that the correlation coefficient between the actual measured photosynthetic rate and the estimated photosynthetic rate by the neural network is 0.95. The photosynthetic rate evaluation using the bioelectric potential response is improved and this correlation coefficient is greater than that analyzed by the neural network using only the illuminating parameters. This result indicates that the information on the plant bioelectric potential response contributed to the accurate estimation of the photosynthetic rate.

**Keywords**: plant bioelectric potential, photosynthetic rate, neural network, plant factory, LED illumination.

## 1. Introduction

The demand for evaluating plant activities has recently grown with the practical realization of artificial agriculture. One of the popular means of artificial agriculture is a plant factory (GOTO, 2012; Kozai, 2013; Oguntoyinbo et al., 2015b). A plant factory controls the cultivation parameters and grows farm products in closed indoor conditions. However, a plant factory entails expensive running costs because of various requirements of the artificial cultivation conditions such as illumination, air conditioning, and irrigation, among others. Therefore, plant activities must be evaluated to optimize the cultivation parameters. Neural networks and deep learning have been used to optimizing the intricate relationship between the cultivation parameters and plant activities following the development of machine learning. Cultivation conditions have been analyzed using a neural network, and the growth rate and crop yield have been estimated (Ehret et al., 2011; Qaddoum et al., 2011). Moreover, future cultivation conditions, such as humidity, irrigation, and $CO_2$ gas fertilizing, have been forecasted using a neural network (He and Ma, 2010; Chen et al., 2019; QIAN et al., 2019).

The photosynthetic rate is an important plant activity because it is directly linked to the growth rate and crop yield. Monitoring the $CO_2$ consumption rate caused by the photosynthetic reaction is a common photosynthetic rate evaluation method. However, a closed chamber or room is necessary to monitor the $CO_2$ consumption. Moreover, this method is not suitable for plant factories that control a higher $CO_2$ concentration to accelerate photosynthesis.

Estimating the photosynthetic rate from the illuminating intensity to plants is also a conventional method. Although the illuminating intensity mainly controls the photosynthetic rate, this method does not reflect the internal activities of plants and other cultivation conditions; therefore, the estimation accuracy is not high. The use of a neural network attempts to estimate the photosynthetic rate accurately from information on the cultivation parameters (Hu et al., 2019).

We focus herein on the plant bioelectric potential to evaluate photosynthesis. The plant bioelectric potential is generated by the concentration difference of the ions inside and outside the plant cells (Bates et al., 1982; Sukhov et al., 2013). The bioelectric potential varies with the changing internal activities of plants, external environments, plant diseases, and stimulus to plants (Intabon et al., 1996; Fromm and Lautner, 2007; Wang and Huang, 2007; Shibata, 2012). Therefore, the measurement and analysis of the bioelectric potential response are expected to be suitable methods to evaluate plant activities in real time without causing injury to plants. The fundamental measurement of the plant bioelectric potential is performed by inserting a micro-needle electrode or a micro-glass electrode into a plant cell; however, in the practical method, attaching disk electrodes on the plant surface is commonly employed. The bioelectric potential response to illumination was also reported by various authors. The fluctuation of the bioelectric potential is induced typically by illuminating plants and stopping the illumination (Uchida et al., 1991; Matsumoto et al., 2000; Spalding, 2000). Moreover, the relationship between the bioelectric potential and the photosynthetic rate has been reported. The amplitude of the potential fluctuation when plants are illuminated is correlated with the photosynthetic rate (Harada, 1999; ANDO et al., 2008). Several studies have attempted to apply the information on the bioelectric potential response to the illumination control in plant factories (Kwon and Lim, 2011; Oguntoyinbo et al., 2015a; Hasegawa et al., 2015). Optimizing the red–blue ratio of LED illumination is necessary in accelerating the photosynthetic rate and increasing the crop yield (GOTO, 2012). However, the relationship between the red–blue ratio and plant activities is intricate, and the suitable ratio depends on the crop species and situations. Therefore, a real-time evaluation of plant activities by monitoring the plant bioelectric potential is one of the solutions to optimize the red–blue ratio (Murohashi et al., 2018).

The amplitude of the bioelectric potential fluctuation responding to the illumination showed a strong correlation with the photosynthetic rate under a fixed experimental condition, except for the illuminating intensity. However, the photosynthetic reaction is well known to be affected not only by the illuminating intensity but also by various ambient conditions. These ambient conditions also intricately affect the bioelectric potential response. Therefore, we consider that the photosynthetic rate is difficult to evaluate using only the amplitude of the potential response under a combination of various cultivation conditions such as ambient $CO_2$ concentration and ambient temperature (Ando et al., 2012, 2013, 2014; Hasegawa et al., 2014). As a solution to this problem, we analyzed the waveform of the bioelectric potential fluctuation through curve fitting and improved the evaluation accuracy of the photosynthetic rate in our previous study (Ando et al., 2014).

We attempted herein to improve the evaluation accuracy of the photosynthetic rate using the plant bioelectric potential response under various combinations of red–green–blue (RGB) ratio and the total intensity of illumination. For this purpose, we employ a neural network to analyze relationship among the plant bioelectric potential, illuminating intensity, and RGB ratio, and estimate the photosynthetic rate accurately. Although not used to evaluate the photosynthetic rate, other studies have reported previously that a neural network was effective in analyzing the plant bioelectric potential (Nambo et al., 2019; Tahyudin and Nambo, 2019). Additionally, the advantages of using a neural network include not only being able to analyze the current plant activities but also being able to forecast and simulate future plant activities if the cultivation parameters are changed or retained.

## 2. Materials and methods

### 2.1 Measurement system of the plant bioelectric potential and photosynthetic rate

In the experiments, we used a foliage plant called golden pothos (*Epipremnum aureum*) to measure the object of the plant bioelectric potential and to evaluate the photosynthetic rate. Golden pothos is tolerant to environmental change, and allows long-term stable experiments. The plant used herein was planted in a small plant pot ($\varphi \times$ height: $10 \times 8$ cm). Five leaves were kept during the experiments. Fig. 1 schematically shows the measurement system of the plant bioelectric potential and the photosynthetic rate. We attached a disk electrode on a leaf surface, where photosynthesis was active, to detect the bioelectric potential response induced by the photosynthetic reactions. As a reference electrode, we attached another disk electrode to a leaf stem, where photosynthesis was inactive. These disk electrodes were silver-silver chloride (Ag/AgCl) electrodes used commonly for electroencephalography (Nihon Kohden, NE-155A). Subsequently, they were glued on the plant with conductive paste (Nihon Kohden, Z-181BE). The potential difference between the two electrodes was measured by a digital multimeter (DMM, 7351E, ADC Corp.). The potential variation was recorded in a personal computer at 1-s sampling intervals.

The plant was put in a clear closed box (AZ ONE, portable desiccator standard PH) which an inner volume of 26.2 L. The photosynthetic rate was evaluated by the $CO_2$ consumption in the closed box caused by photosynthesis of the plant using a $CO_2$ sensor (PP systems, SBA-4). The temperature and humidity in the closed box were kept constant from 25 to 27 °C and from 37 to 47%, respectively. Plant watering was performed before the experiments.

The experiments were conducted in a dark room. An LED panel (Nippon Medical & Chemical Instrument, 3LH-256) was used as the light source. The LED panel consisted of blue (445 nm), green (520 nm), and red (660 nm) LEDs. The illuminating intensity was controlled by a current control module that we made. The illuminating intensity was measured by a spectral irradiance meter (Konica Minolta, CL-500A) as the photosynthetic photon flux density. In the case of illuminating the LEDs with full power, the total intensity of RGB LEDs was 325 μmol/m²·s with 9.5 cm distance between the LED panel and the leaf attached with the electrode. The red, green, and blue intensities were 130, 65, and 130 μmol/m²·s with this distance, respectively. Hereinafter, the illuminating parameters are referred to as the total intensity and each RGB intensity. We measured the bioelectric potential and the $CO_2$ consumption of the plant under various combinations of the total intensity and the RGB ratio. Additionally, the plant was kept in a dark condition for 1 h before starting the illumination to adjust the experimental condition.

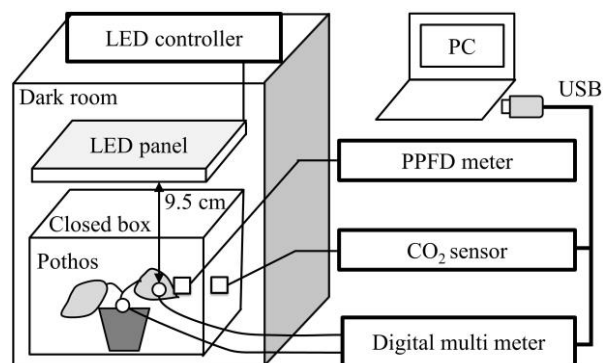

Fig. 1. Measurement system of the plant bioelectric potential and the $CO_2$ consumption of the plant.

Fig. 2 shows a typical example of the bioelectric response to illumination. After starting the illumination, the bioelectric potential fluctuated, then recovered and stayed around its previous potential before the illumination started. We defined the difference between the bottom and the peak of the potential fluctuation as amplitude $a$ and the time from the bottom to the peak as response time $t$ (Fig. 2). The amplitude $a$ and the gradient $a/t$ were used as the parameters of the bioelectric response in the neural network analysis.

Fig. 3 shows an example of the $CO_2$ concentration change in the closed box with the photosynthetic reaction of the plant. In the illumination period, the $CO_2$ consumption rate of the plant increased gradually, then became constant. Therefore, we used the $CO_2$ concentration at 20 and 30 min after starting the illumination and evaluated the $CO_2$ consumption rate for 10 min. The $CO_2$ consumption rate by the plant in the closed box was defined as the "measured" photosynthetic rate.

**2.2 Structure of the neural network and data sets**

Fig. 4 shows the structure of the neural network used to estimate the photosynthetic rate by analyzing the bioelectric potential response and the illuminating parameters. This neural network had three layers: input, hidden, and output layers. The units in the hidden and output layers had a threshold function $\theta$. We used the hyperbolic tangent as the threshold function. The number of units $n$ in the hidden layer was variable in each investigation. Each node connecting the units had a weight parameter $\omega$. These weight parameters were updated by back propagation learning. The measured photosynthetic rate obtained by the $CO_2$ sensor was used as the teaching data. Thus, this neural network output the "estimated" photosynthetic rate.

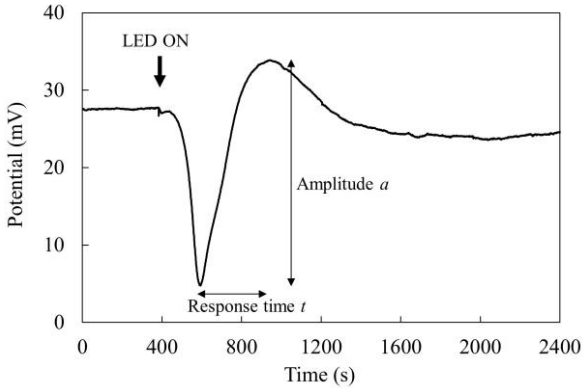

Fig. 2. Example of the plant bioelectric potential response to illumination.

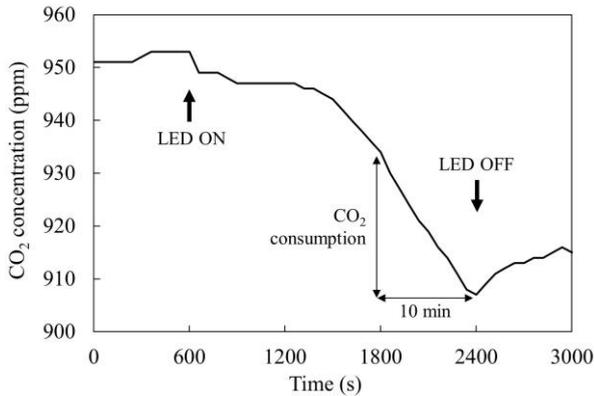

Fig. 3. Example of the $CO_2$ concentration change and definition of the measured photosynthetic rate.

We used the following six input data: amplitude $a$ and gradient $a/t$ of the bioelectric potential response, the total intensity of illumination, and each RGB intensity. We used the 12 data sets in Table 1. Each data set consisted of 6 units of input data and the measured photosynthetic rate as the teaching data. In addition, all data were normalized from −1 to 1 for input in the neural network.

A training phase of the neural network was performed with 11 data sets. After sufficient training times, a data set out of use in the training phase was input to the neural network as an unknown data set. The output value with the unknown data set corresponds to the estimated photosynthetic rate. We then compared the estimated photosynthetic rate with the measured photosynthetic rate to evaluate the estimation accuracy. The closer these values, the more accurate the estimation of the photosynthetic rate. We changed the combination of 11 data sets for training and the unknown data set for estimation in 12 patterns. Finally, we obtained the correlation coefficient and the residual sum of squares between the measured and estimated photosynthetic rates.

**3. Results and discussion**

**3.1 Correlation between each input datum and the measured photosynthetic rate**

Before analyzing using the neural network, we first checked the correlation coefficient between the measured photosynthetic rate and each input datum as a control shown in Fig. 5. The total intensity and the red intensity of illumination had higher correlation coefficients because illuminating caused the photosynthetic reaction, and the red wavelength was the most effective in photosynthetic reaction. Fig. 6 shows the correlation between the total intensity and the measured photosynthetic rate. The correlation coefficient was 0.82. Even though the illuminating intensity was used as a conventional estimation method of the photosynthetic rate, this correlation coefficient was not sufficient to evaluate the photosynthetic rate in the practical cultivation scene.

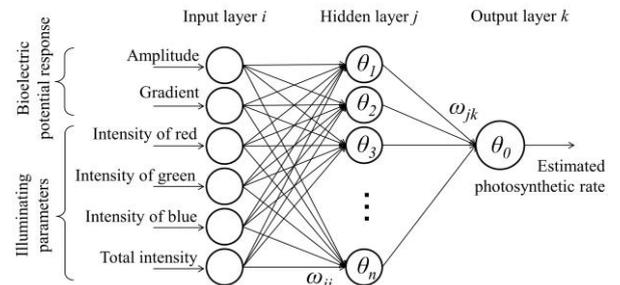

Fig. 4. Structure of the neural network used to estimate the photosynthetic rate.

Table 1. Data set for inputting the neural network and teaching data.

| Data set | Amplitude $a$ (mV) | Gradient $a/t$ (mV) | Illuminating intensity (µmol/m²·s) | | | | Measured photosynthetic rate (ppm/min) |
| --- | --- | --- | --- | --- | --- | --- | --- |
| | | | Red | Green | Blue | Total | |
| #1 | 8.05 | 0.0095 | 130 | 65 | 130 | 325 | 2.7 |
| #2 | 6.72 | 0.0096 | 98 | 48 | 98 | 244 | 2.5 |
| #3 | 8.11 | 0.0035 | 65 | 33 | 65 | 163 | 2.3 |
| #4 | 4.26 | 0.0040 | 32 | 16 | 33 | 81 | 0.9 |
| #5 | 1.09 | 0.0406 | 130 | 0 | 130 | 260 | 2.8 |
| #6 | 7.87 | 0.0089 | 130 | 65 | 0 | 195 | 2.8 |
| #7 | 13.67 | 0.0316 | 0 | 65 | 130 | 195 | 1.7 |
| #8 | 12.07 | 0.0093 | 130 | 0 | 0 | 130 | 2.2 |
| #9 | 7.87 | 0.0075 | 98 | 0 | 0 | 98 | 1.2 |
| #10 | 5.41 | 0.0079 | 0 | 65 | 0 | 65 | 1.3 |
| #11 | 13.70 | 0.0077 | 0 | 49 | 0 | 49 | 1.0 |
| #12 | 9.11 | 0.0038 | 0 | 0 | 130 | 130 | 1.6 |

In contrast, even though the photosynthetic rate was reportedly evaluated by the potential fluctuation amplitude (Harada, 1999; ANDO et al., 2008), the amplitude had no correlation, as shown in Fig. 7. We considered that the relationship between the amplitude and the photosynthetic rate was changed because of different illuminating RGB ratio.

### 3.2 Result of analyzing the bioelectric potential and illuminating parameters by the neural network

This section presents the result of analyzing the bioelectric potential response and the illuminating parameters by the neural network. The number of units in the hidden layer as well as the number of input data was 6. The number of training times was a power of 2 up to $2^{13}$. The learning coefficient was 0.1 and the momentum was 0.04. Fig. 8 shows the correlation coefficient and the residual sum of squares between the measured and estimated photosynthetic rates. At each training time in Fig. 8, these values were averaged and obtained from 10 iteration times. The error bars indicate each standard deviation (SD). The correlation coefficient, residual sum of squares, and their SDs improved with increasing the number of training times. At 1024 training times, the residual sum of squares showed a minimum value of 0.91, while the correlation coefficient showed a maximum value of 0.95. After 1024 training times, these values became worse because of over training. Although the number of units in the hidden layer increased, this trend was almost the same as that in the case of 6 units.

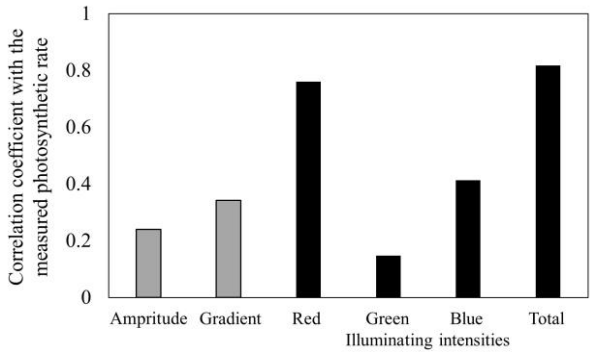

Fig. 5. Correlation coefficient between each input datum and the measured photosynthetic rate.

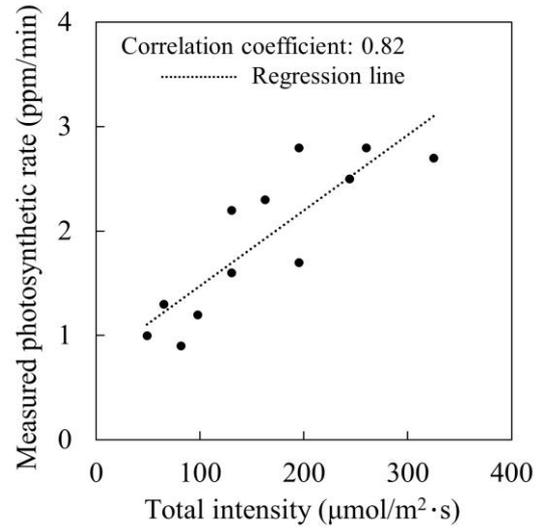

Fig. 6. Correlation between the total intensity and the measured photosynthetic rate.

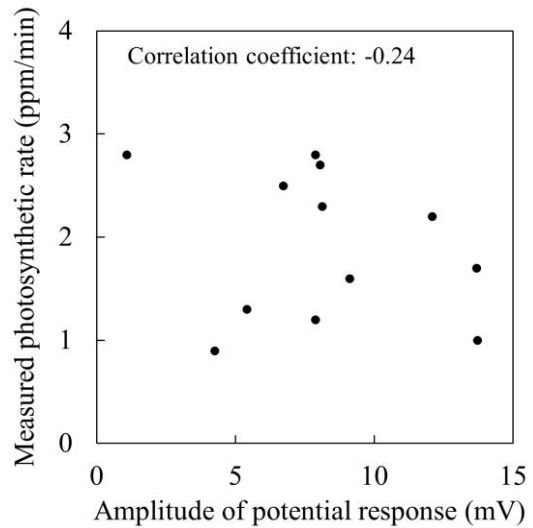

Fig. 7. Correlation between the amplitude of the bioelectric potential fluctuation and the measured photosynthetic rate.

Fig. 9 shows the correlation between the measured and estimated photosynthetic rates at 1024 training times. Each dot was obtained from an average of 10 iteration times. The error bars indicate the SDs. The numbers near each dot correspond to the data set number in Table 1. The correlation was greatly improved compared with the correlation using the raw amplitude of the bioelectric potential fluctuation in Fig. 7. This 0.95 correlation coefficient was better than that with the total intensity of illumination shown in Fig. 6. Focusing on each estimation, even though #3, #7, #8, and #12 had large SDs, they had no common tendency causing the variation of the estimation in their data sets. #2 and #10 had large estimation errors. However, they also did not have a common tendency in their data sets. We considered the absence of a data set similar to another data as a reason for these large SDs and errors. Nevertheless, the neural network analyzed the relationship between the bioelectric potential response and the illuminating parameters, and estimated the photosynthetic rate accurately.

Additionally, when all 12 data sets were used in the training phase, and the neural network estimated the photosynthetic rate by a known input data set included in the 12 training data sets, the estimation was perfect, and the correlation coefficient between the measured and estimated photosynthetic rates was 1.000. This result suggested that the estimation accuracy of the photosynthetic rate was improved with the training of a large number of data sets containing similar data sets. Finally, Fig. 10 shows the correlation between the measured and estimated photosynthetic rates by the neural network using only the illuminating parameters "total intensity" and "RGB intensity." The SDs were small and the error bars were hidden inside each dot. Its correlation coefficient was 0.88, which improved compared with that using the total intensity in Fig. 6. However, the correlation coefficient analyzed with both the bioelectric potential response and the illuminating parameters was much better than that. This result indicated that the information on the plant bioelectric potential response contributed to the accurate estimation of the photosynthetic rate.

## 4. Conclusions

In this study, we attempted to improve the evaluation accuracy of the photosynthetic rate using the plant bioelectric potential response and employed a neural network to analyze the relationship among the plant bioelectric potential, illuminating intensity, RGB ratio, and photosynthetic rate. Even though various combinations of the RGB ratio of illumination affected the bioelectric potential response and the photosynthetic rate, we succeeded in the accurate estimation of the photosynthetic rate by analyzing the relationship between the bioelectric potential response and the illuminating parameters using the neural network. The obtained correlation coefficient between the actual measured and estimated photosynthetic rates showed 0.95. This correlation coefficient was greater than that analyzed by the neural network using only the illuminating parameters. This result indicated that the information on plant bioelectric potential response contributed to the accurate estimation of the photosynthetic rate.

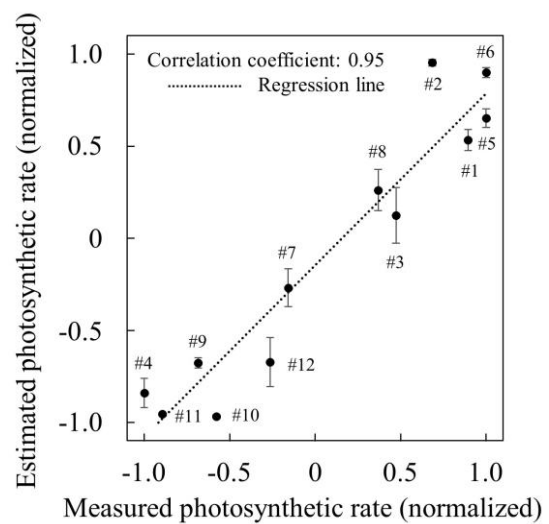

Fig. 9. Correlation between the measured and estimated photosynthetic rates analyzed with the bioelectric potential response and the illuminating parameters.

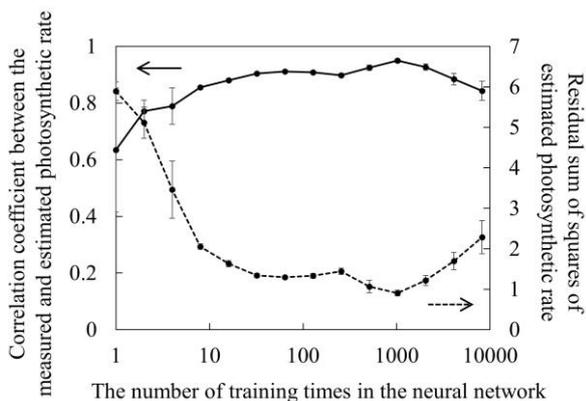

Fig. 8. Correlation coefficient and residual sum of squares between the measured and estimated photosynthetic rates analyzed with the bioelectric potential response and the illuminating parameters.

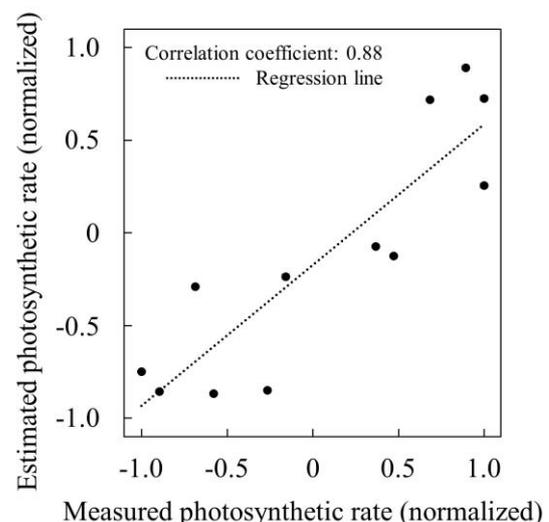

Fig. 10. Correlation between the measured and estimated photosynthetic rates analyzed with only illuminating parameters.

In future work, we must refine the structure of the neural network specialized to estimate the photosynthetic rate as a deep learning. In addition, accumulating more data sets for neural network learning and increasing the number of measurement parameters related to photosynthesis are necessary for a better estimation of the photosynthetic rate.

Using a neural network has distinct advantages (i.e., not only being able to analyze current plant activities but also being able to forecast and simulate future plant activities). Meanwhile, the plant bioelectric potential is available to evaluate internal plant activities and diseases as well as photosynthesis in real time without injury to plants. Therefore, the combination of the neural network and the plant bioelectric potential is expected to greatly contribute to the optimization of the cultivation parameters in plant factories.

**Acknowledgment**

This work was supported by JSPS KAKENHI Grant Number JP16K18116.

**References**

ANDO, K., HASEGAWA, Y., MAEKAWA, H., KATSUBE, T., 2008. Analyzing Bioelectric Potential Response of Plants Related to Photosynthesis under Blinking Irradiation. IEICE Trans. Electron. 91, 1905–1910. https://doi.org/10.1093/ietele/e91-c.12.1905

Ando, K., Hasegawa, Y., Uchida, H., Kanasugi, A., 2014. Analysis of plant bioelectric potential response to illumination by curve fitting. Sensors Mater. 26, 471–482.

Ando, K., Hasegawa, Y., Yaji, T., Uchida, H., 2013. Study of plant bioelectric potential response due to photochemical reaction and carbon-fixation reaction in photosynthetic process. Electron. Commun. Japan 96, 85–92. https://doi.org/10.1002/ecj.11518

Ando, K., Hasegawa, Y., Yaji, T., Uchida, H., 2012. Study of plant bioelectric potential response due to photosynthesis reaction. Electron. Commun. Japan 95, 10–16. https://doi.org/10.1002/ecj.11393

Bates, G.W., Goldsmith, M.H.M., Goldsmith, T.H., 1982. Separation of tonoplast and plasma membrane potential and resistance in cells of oat coleoptiles. J. Membr. Biol. 66, 15–23. https://doi.org/10.1007/BF01868478

Chen, Z., Zhao, C., Wu, H., Miao, Y., 2019. A Water-saving Irrigation Decision-making Model for Greenhouse Tomatoes based on Genetic Optimization T-S Fuzzy Neural Network. KSII Trans. Internet Inf. Syst. 13, 2925–2948. https://doi.org/10.3837/tiis.2019.06.009

Ehret, D.L., Hill, B.D., Helmer, T., Edwards, D.R., 2011. Neural network modeling of greenhouse tomato yield, growth and water use from automated crop monitoring data. Comput. Electron. Agric. 79, 82–89. https://doi.org/10.1016/j.compag.2011.07.013

Fromm, J., Lautner, S., 2007. Electrical signals and their physiological significance in plants. Plant, Cell Environ. 30, 249–257. https://doi.org/10.1111/j.1365-3040.2006.01614.x

GOTO, E., 2012. Plant production in a closed plant factory with artificial lighting. Acta Hort 956, 37–49.

Harada, K., 1999. Effects of Light Intensity and Abscisic Acid on the Light-induced Changes of Electric Potential in Eggplants. IEEJ Trans. Sensors Micromachines 119, 270–278. https://doi.org/10.1541/ieejsmas.119.270

Hasegawa, Y., Hoshino, R., Uchida, H., 2015. Development of Cultivation Environment Control System Using Plant Bioelectric Potential, in: Proceedings of NOLTA. pp. 860–863.

Hasegawa, Y., Yamanaka, G., Ando, K., Uchida, H., 2014. Ambient temperature effects on evaluation of plant physiological activity using plant bioelectric potential. Sensors Mater. 26, 461–470.

He, F., Ma, C., 2010. Modeling greenhouse air humidity by means of artificial neural network and principal component analysis. Comput. Electron. Agric. 71. https://doi.org/10.1016/j.compag.2009.07.011

Hu, J., Xin, P., Zhang, S., Zhang, H., He, D., 2019. Model for tomato photosynthetic rate based on neural network with genetic algorithm. Int. J. Agric. Biol. Eng. 12, 179–185. https://doi.org/10.25165/j.ijabe.20191201.3127

Intabon, K., Maekawa, T., Sato, K., 1996. Studies on the monitoring techniques and application of bioelectric potentials in bioproduction, 1: Relationship between foliar temperature and electric potential of foliar surface in a tomato plant [Lycopersicon esculentum]. J. Soc. Agric. Struct. 33–40.

Kozai, T., 2013. Resource use efficiency of closed plant production system with artificial light: Concept, estimation and application to plant factory. Proc. Japan Acad. Ser. B Phys. Biol. Sci. 89, 447–461. https://doi.org/10.2183/pjab.89.447

Kwon, S., Lim, J., 2011. Improvement of Energy Efficiency in Plant Factories through the Measurement of Plant Bioelectrical Potential BT - Informatics in Control, Automation and Robotics,

Matsumoto, K., Shibusawa, S., Sakai, K., Sasao, A., 2000. Measuring the Leaf Electricity of Living Plant. IFAC Proc. Vol. 33, 135–139. https://doi.org/https://doi.org/10.1016/S1474-6670(17)36765-4

Murohashi, F., Uchida, H., Hasegawa, Y., 2018. Evaluation of photosynthetic activity by bioelectric potential for optimizing wavelength ratio of plant cultivation light 281–287. https://doi.org/10.15406/ijbsbe.2018.04.00141

Nambo, H., Tahyudin, I., Nakano, T., Yamada, T., 2019. Comparison of deep learing algorithms for indoor monitoring using bioelectric potential of living plants. Proc. - 2018 3rd Int. Conf. Inf. Technol. Inf. Syst. Electr. Eng. ICITISEE 2018 110–113. https://doi.org/10.1109/ICITISEE.2018.8720992

Oguntoyinbo, B., Hirama, J., Yanagibashi, H., Matsui, Y., Ozawa, T., Kurahashi, A., Shimoda, T., Taniguchi, S., Nishibori, K., 2015a. Development of the SMA (speaking mushroom approach) environmental control system: Automated cultivation control system characterization. Environ. Control Biol. 53, 55–62. https://doi.org/10.2525/ecb.53.55

Oguntoyinbo, B., Saka, M., Unemura, Y., Hirama, J., 2015b. Plant factory system construction: Cultivation environment profile optimization. Environ. Control Biol. 53, 77–83. https://doi.org/10.2525/ecb.53.77

Qaddoum, K., Hines, E., Illiescu, D., 2011. Adaptive neuro-fuzzy modeling for crop yield prediction. Recent Res. Artif. Intell. Knowl. Eng. Data Bases - 10th WSEAS Int. Conf. Artif. Intell. Knowl. Eng. Data Bases, AIKED'11 199–204.

QIAN, W., LIU, Y., LI, Y., SHEN, T., XU, T., WANG, X., XIE, W., LI, C., SUN, H., 2019. Study on Strawberry $CO_2$ Gas Fertilizer in Greenhouse Based on BP Neural Network. DEStech Trans. Comput. Sci. Eng. 152–158. https://doi.org/10.12783/dtcse/iteee2019/28735

Shibata, S., 2012. Relation Between Characteristics of Plant Bioelectric Potential and Purification Function Under LED Light, in: Kimura, H. (Ed.), . IntechOpen, Rijeka, p. Ch. 15. https://doi.org/10.5772/26957

Spalding, E.P., 2000. Ion channels and the transduction of light signals. Plant. Cell Environ. 23, 665–674. https://doi.org/10.1046/j.1365-3040.2000.00594.x

Sukhov, V., Akinchits, E., Katicheva, L., Vodeneev, V., 2013. Simulation of Variation Potential in Higher Plant Cells. J. Membr. Biol. 246, 287–296. https://doi.org/10.1007/s00232-013-9529-8

Tahyudin, I., Nambo, H., 2019. Comparison study of deep learning and time series for bioelectric potential analysis. Proc. - 2018 3rd Int. Conf. Inf. Technol. Inf. Syst. Electr. Eng. ICITISEE 2018 79–83. https://doi.org/10.1109/ICITISEE.2018.8720998

Uchida, T., Nakanishi, Y., Sakano, T., 1991. Measurement of bioelectric potential on the surface of Spinach lamina. IFAC Proc. Vol. 24, 373–378. https://doi.org/10.1016/b978-0-08-041273-3.50069-0

Wang, C., Huang, L., 2007. Monitoring and analysis of electrical signals in water‐stressed plants. New Zeal. J. Agric. Res. - N Z J AGR RES 50, 823–829. https://doi.org/10.1080/00288230709510356

in: Tan, H. (Ed.), . Springer Berlin Heidelberg, Berlin, Heidelberg, pp. 641–648.